\documentclass{article}
\usepackage[utf8x]{inputenc}
\usepackage{amsmath, amssymb}
\usepackage{tikz}
\usepackage{subcaption}

\newcommand{\mbf}[1]{\mathbf #1}

\def\x{\mbf x}

\def \k {\mbf k}

\def\dph{{\delta\hat\phi}}
\def\delphi{{\delta\phi}}

\title{Multipoint correlators in multifield cosmology}
\author{George Panagopoulos and Eva Silverstein}

\begin{document}

\maketitle

\begin{abstract}
    Connected $N$-point amplitudes in quantum field theory are enhanced by a factor of $N!$ in appropriate regimes of kinematics and couplings, but the non-perturbative analysis of this for collider physics applications is subtle. We resolve this question for $N$-point correlation functions of cosmological perturbations in multifield inflation, and comment on its application to primordial non-Gaussianity. We find that they are calculably $N!$-enhanced using a simple model for the mixing of the field sectors which leads to a convolution of their probability distributions.  This effect leads to model-dependent but interesting prospects for enhanced observational sensitivity.
    
\end{abstract}

\newpage

\tableofcontents

\section{Introduction}
The behavior of multi-point correlation functions and S-matrix amplitudes at large particle number is of interest for various applications.
At tree-level, there is an $N!$ enhancement of large $N$ $N$-point correlation functions in perturbative quantum field theory as initially studied by Voloshin in \cite{Voloshin} and developed by many authors \cite{Browntree} \cite{Argyresetal} \cite{Sonetal} \cite{Khozeetal} \cite{RajuN}.  

For S-matrix amplitudes that produce $N$ outgoing quanta, this occurs because the contributions from low-order interaction vertices build up many tree diagrams, of order $N!$. For some regimes of couplings and kinematics, this enhancement is known to survive the sum over tree diagrams (which can be derived equivalently from the classical field configuration)  and to persist in the presence of sufficiently small quantum corrections.  The interaction probability -- obtained by squaring the amplitude and integrating over final particle momenta -- contains one $N!$ in the denominator in the phase space for identical particles, leaving a net enhancement.  For example, in $\lambda\phi^4$ theory, the $1\to N$ amplitude near threshold is of order $\lambda^{N/2} N! + \text{loops}$, and the decay probability is of order $\lambda^N N! + \text{loops}$.  

In the setting of particle decays and scattering it is not clear to what extent this effect survives in the quantum theory when $\lambda N$ is not perturbatively small.  As noted in \cite{Khozeetal}, if it did persist it would have dramatic implications for Higgs physics, leading to a large decay width for the Higgs: the Higgs would fail to be a good quasiparticle at a relatively low energy scale.  More recent analyses \cite{criticalrefs}\ do not obtain such growth in somewhat similar quantities; still, by investigating this they uncover an interesting emergent 'tHooft expansion arising from a semiclassical approximation, related to results in large-charge quantum field theory \cite{largecharge}.  From this perspective, it seems interesting in contrast that we will find a positive result for factorial growth in the setting of early universe cosmology, where the required calculations are actually easier to control.  

In time-dependent backgrounds, such as that arising in early universe cosmology, we may ask a similar question for $0\to N$ processes.  A prime example is the set of connected $N$-point in-in correlation functions relevant for studies of primordial non-Gaussianity, the moments of the probability distribution for scalar fluctuations.
The main object of interest there is the wavefuntion of primordial perturbations which seed the structure in the universe. We may write it schematically as
\begin{equation}
    \Psi(\zeta(\mathbf{x}), \gamma(\mathbf{x}), \{\chi(\mathbf{x}) \}; \{ \lambda \})
\end{equation}
where $\zeta$ and $\gamma$ are the scalar and tensor perturbations, $\{ \chi \}$ represents additional sectors of fields not directly observable, and $\{ \lambda \}$ denotes the parameters (couplings) of the theory that generates the perturbations. The probability distribution for the observables $\zeta$, $\gamma$ is derived from this by tracing over the $\chi$ sector,
\begin{equation}\label{probdist}
    {\cal L}(\zeta(\x), \gamma(\x)| \{\lambda\})=\text{Tr}[\rho |\zeta\gamma\rangle \langle \zeta \gamma|] =\int D\chi |\Psi |^2, ~~~~\rho=\text{Tr}_\chi[|\Psi\rangle\langle\Psi|].
\end{equation}

Observations indicate that this is at least approximately Gaussian \cite{Gdata}.   A Gaussian distribution arises in free field theory { when} the system starts in its ground state (or any other Gaussian initial state).  In any other situation, the state is non-Gaussian at some level.  For a perturbative quantum field theory, the non-Gaussianity vanishes in the limit of zero couplings $\{\lambda\}\to 0$.  But for mildly perturbative couplings (such as those arising in particle physics at appropriate scales, with $\lambda\sim 10^{-2}$), the effects of interactions are not arbitrarily small and it is interesting to compute their effects and constrain them with data as systematically as possible.  

%Previous analyses of non-Gaussianity have largely focused on low moments of this distribution, with the idea that those would always dominate the signal to noise ratio for perturbative quantum field theory.  We will see that this assumption is not correct for an interesting range of perturbative
%couplings, leading to a new window of opportunity to empirically discover or better constrain the relevant parameters $\{\lambda\}$.  

In situations where the quantum fields in the early universe interact arbitrarily weakly, one can immediately characterize this via low-point correlation functions.  These are already rich with different possible shapes in kinematic space \cite{shapes, Gdata}\ which encode various aspects of the dynamics.  
However, even within the class of field theories with perturbative couplings $\lambda <1$, interaction effects can build up during inflation \cite{Starobinsky}\cite{GS}\cite{productive}\ and reheating \cite{BondBilliards}.  This in turn can lead to non-Gaussianity that is not well captured by the lowest-point correlation function \cite{productive}\cite{BondBilliards}\cite{AndreiWeb}\cite{PBHpaper}\cite{PajerN}\cite{Bruno}.\footnote{See also \cite{outsideHNG}\ for an interesting analysis of multifield evolution beyond the observable horizon and its relation to multipoint correlators and local inferences.}     

The structure of the paper is as follows: After explaining qualitatively why the $N!$ enhancement is tractable in dS space in Section 2, we present the methods in detail in section 3. We then present the enhancement for a toy theory that is fully solvable, and prove it for a large class of theories in section 4.
In section 5, we investigate implications of this for primordial non-Gaussianity searches, and in section 6 we summarize and mention directions for further research.   

\section{Simplifications of dS space and local\\non-Gaussianity}
It is perhaps surprising that we are able to derive a general $N!$ enhancement for correlation functions in an inflationary setting while no such result exists for Minkowski space. 
The results on Minkowski space are accessible in specific regimes of coupling and kinematics in the theory of interest. For example, \cite{Voloshin} focuses on $\lambda \phi^4$ and \cite{Sonetal} on the weak-coupling multi-particle limit $\lambda n\to \epsilon$, with $\lambda$ being the coupling, $n$ the number of particles produced and $\epsilon$ fixed.

The simplicity of dS space comes in the freezing out of the modes after horizon crossing. In the multifield context, there remains meaningful dynamics outside the horizon, and the dilution of gradients enables the stochastic approach to inflation \cite{SalopekBond}\cite{Starobinsky}\  which descends from the full quantum theory as in \cite{GS}. Those approaches are able to resum some of the loop contributions by exploiting the fact that $-k\eta \ll 1$, where $\eta\sim - e^{-Ht}/H$ is the proper time which decays at late time exponentially in FRW time $t$. In QFT in Minkowski space, one has much less control over the loop effects that could spoil the tree-level enhancement of the correlation functions. That is why specific regimes of the phase-space were enforced by hand in the initial investigations of the flat spacetime problem.  In the cosmological case, the accelerated expansion itself restricts the phase space naturally.

This is not the first time this phenomenon of a greater simplicity in de Sitter than in flat spacetime has arisen.  It has even made an appearance in rigorous mathematics (related to physics):  the proof of stability of Kerr black holes \cite{Kerrstability}\ pertains in de Sitter spacetime but not otherwise.  This is for a similar reason, involving the dilution of perturbations from the accelerated expansion.

The multifield inflationary scenarios that generate local non-Gaussianity captures this simplicity. It enables us to analyze large tails of the primordial scalar perturbations in a controlled way \cite{PBHpaper}. After the exit from inflation and all the long modes are frozen out, we mix the additional field sectors with the inflaton. As we will see, in the mixing, dS helps us again by suppressing the momentum conjugate to the inflaton by $a^3$ and enabling us to write the wavefunction evolved by the mixing Hamiltonian as a simple shift in field space.

It would be interesting to explore whether single-field inflationary perturbations can produce the same N!-enhanced correlation functions that we find here\footnote{This goes beyond the low-point functions analyzed in e.g. \cite{Juan}\cite{DBI}\cite{EFT}) following early work including \cite{SalopekBond}\cite{EarlyNG}.}. There, correlation functions would go like (assuming the tree diagrams constructively add up)
\begin{equation}
    \langle \zeta_1 \cdots \zeta_n \rangle \sim N! \lambda^{\alpha N} (1+c_1 \lambda^{\beta} N^2+\dots)
\end{equation}
with $\alpha$ and $\beta$ being constants depending on the order of the interaction and $\lambda$ a dimensionless coupling constant. For example, for a cubic interaction, $\alpha = 1$ and $\beta = 2$, and for a quartic interaction, $\alpha = \frac{1}{2}$ and $\beta = 1$. The leading loop effects come from joining any 2 lines with a propagator, and any two lines at a point respectively. These loop effects can be very large and require resumming. For $\lambda \phi^4$ in some regimes, previews work \cite{Sonetal} was able to resum the contributions controlled by $\lambda N^2$, relegating the question to the effect of those controlled by $\lambda N$.  Even those may be calculable, although this case seems more similar to the particle physics case (\cite{criticalrefs}\ versus \cite{Khozeetal}), something that would be interesting to generalize to cosmological correlators. We will leave this to future work.

\section{General Setup and Methods}
Observations of cosmological scalar\footnote{From now on we suppress the tensor perturbations, which have not been detected at least as of this writing. However, our analysis can be straightforwardly generalized to include tensor modes.} perturbations $\zeta(\x)$ may be compared to those predicted by a theoretical probability distribution depending
on some parameters $\{\lambda \}$.   The likelihood, or probability of the data given the theory, is given by squaring and tracing over the non-observable fields as in (\ref{probdist}):
\begin{equation}\label{Likgen}
{\cal L}(\zeta(\x) |\{\lambda\})= \int D\chi |\Psi(\zeta(\x), \chi(\x); \{\lambda\})|^2.
\end{equation}

%We can write this in terms of the conjugate momentum to the inflaton fluctuation $\delta\phi$, $\Pi_{\delta\phi} = a(t)^3 \delta \dot{\phi}$

Ideally we would compute this functional theoretically, and compare it to data directly.  At CMB scales, we would evaluate it on the map; large scale structure may enable a volume's worth of data points, and in \cite{PBHpaper}\ we were led to shorter-scale probes.  This determines whether, according to the theory,  the data is higher-probability with null values $\{\lambda\}=0$ or for some nonzero values of the couplings (and with what significance).   That is not always tractable, so it is useful to work with other quantities derived from the full likelihood.

%One diagnostic of the information available to distinguish the Non-Gaussian probability distribution from the Gaussian one is the relative entropy (also known as the Kullback–-Leibler divergence), an average of the log of the ratio of likelihoods:
% \begin{equation}\label{RelSdef}
% {\cal S}_{rel}\equiv \int D\zeta P(\zeta)\log\left(\frac{{\cal L}(\zeta(\x) |\{\lambda\})}{{\cal L}(\zeta(\x) |\{0\})}\right)
% \end{equation}
% Here, $P(\zeta)$ may be taken to be either of the two probability distributions; ${\cal S}_{rel}$ is not symmetric.  The first term in its Taylor expansion 
% about $\{\lambda\}=0$ is the Fisher metric $F_{\lambda\lambda}$.  As we will see, in some cases, the relative entropy (\ref{RelSdef}) is well approximated by the first term in the Taylor expansion,  the constraint on $\lambda$ is well estimated by the inverse of the Fisher metric, and low-point correlation functions suffice to achieve this constraint.  In other cases, this first term is subdominant, and there is more information available (e.g. on the tail of the distribution).   Moreover, certain observables (such as constraints on PBHs and mechanisms for their production) are specifically sensitive to the tail.   

The set of connected correlation functions of $\zeta$ is a useful quantity, which is sometimes easier to compute than the full probability distribution.  These are generated by $W(J)$, defined by
\begin{equation}\label{WJ}
e^{W(J)}= \int D\zeta e^{\int J\zeta} {\cal L}(\zeta | \{\lambda\})
\end{equation}
by taking $N$ functional derivatives with respect to $J$:
\begin{equation}\label{NpfW}
\langle \zeta_{\k_1}\dots\zeta_{\k_N}\rangle_C = { \left.\frac{\delta^N}{\delta J_{\k_1}\dots \delta J_{\k_N}} W(J)\right|_{J_{\k_i}=0}}
\end{equation}
setting $J$ to zero at the end.
We will find that these connected correlators scale like $N!$ in a wide class of inflationary scenarios with at least one additional light field.  In some cases, there is also an exponential enhancement $\sim \lambda_r^N$, with $\lambda_r$ a ratio of couplings in the model.  
%One of our main goals is to understand the phenomenological implications of this factorial growth in the universality class of models exhibiting it, and how it depends on model parameters.

To simplify the analysis, we will often work with another quantity derived from the full likelihood -- the histogram of temperature fluctuations, also known as the one-point probability density function.  
Given a realization of the field, we can count points with a given fluctuation $\hat\zeta$: 
\begin{equation}\label{histfromP}
N_{\hat\zeta}= k_{\rm max}^3 \int d\x' \delta(\zeta(\x')-\hat\zeta)
%N_{\hat\zeta}= \int d\x' H \delta(\zeta(\x')-\hat\zeta)
\end{equation}
where $1/k_{\rm max}$ is the resolution of the survey, which for simplicity is assumed to be uniform. We can compare this to the average of the histogram according to the field-theoretic distribution (\ref{Likgen}), given by
\begin{equation}\label{histfromPavg}
\langle N_{\hat\zeta}\rangle=k_{\rm max}^3 \int d\x' \int D\delta\zeta(\x) {\cal L}(\zeta(\x)|\{\lambda\}) \delta(\zeta(\x')-\hat\zeta)
%\langle N_{\hat\zeta}\rangle= \int d\x' \int D\delta\zeta(\x) {\cal L}(\zeta(\x)|\{\lambda\}) H \delta(\zeta(\x')-\hat\zeta)
\end{equation}
This is the probability of measuring a given value of $\zeta$, $\hat\zeta$ at one point, having traced out the field at other points.  It can also be used to calculate the N point functions at a single point.   In scenarios containing one or more additional light non-shift-symmetric fields present during inflation, this theoretical averaged histogram is determined by a combination of the stochastic methods of Starobinsky as in \cite{Starobinsky}, and the mixing between field sectors. In different regimes one or the other of these may be relevant.  We will review this and make use of it below.

\subsection{Local non-Gaussianity}\label{twofield}     

%\subsection{Analysis in two-field examples using outside-horizon ultralocality}\label{subsec:twofield}
A standard form of non-Gaussianity with amplitude parameterized by $f_{\rm NL}^{\rm local}$ is sensitive to the presence of one or more additional fields $\chi$.  If these are light, they develop a variance during inflation similar to that of the inflaton perturbations $\delta\phi$. But unlike the inflaton field, their super-horizon interactions are not constrained by symmetries,
%as constrained by the (generalized) slow roll conditions $\dot H/H^2, \ddot H/H^3 \ll 1$, 
and they may imprint nonlinearities on the scalar perturbations via a variety of mechanisms \cite{BondBilliards}\cite{LocalNGmechanisms}.  Their evolution outside the horizon is ultralocal, as we will review shortly.  At the level of the bispectrum, the local shape of non-Gaussianity, which contains correlations between long and short modes, can only be generated if at least one additional field is present \cite{Juan}\cite{EFT}.  

In this section, we will set up a class of models of this kind and determine the relative importance of the bispectrum versus other aspects of the distribution, including the power spectrum and higher point correlators.
We will make some special choices in specifying the scenario in order to make the calculations as simple as possible.  After deriving the factorial enhancement explicitly in a simple example, we will show that it extends to a much wider class of models.  
%We will then study in detail a 2-parameter subpace of models, determining under what circumstances the low-point functions do not capture the leading non-Gaussianity.   

Consider a system with two fields, the inflaton $\phi$ and another scalar $\chi$.   We denote the wave functional of the perturbations $\delta\phi(\x)$ and $\chi(\x)$ as $\Psi(\delta\phi(\x), \chi(\x), t)$; we will eventually trace out $\chi$ because {  $\zeta\sim H\delta\phi/\dot\phi$ will be the directly observed scalar perturbation}. {We are interested { for simplicity} in cases where the scalar perturbation is dominated by the mostly-Gaussian fluctuations of $\delta\phi$, but where there is an additional, potentially very-non-Gaussian contribution, which will dominate in higher-$N$ $N$-point functions of $\delta\phi$.} {This is somewhat analogous to the cases in \cite{productive}, although the origin of the enhanced non-Gaussianity will be different (coming from factorial enhancements of connected correlation functions).}  As above, the probability distribution at  time $t_0$ will be given by the functional integral
\begin{equation}\label{Probdelphi}
P(\delta\phi)=\int D\chi |\Psi(\delta\phi, \chi, t_0)|^2 = \text{Tr} [\rho |\delta\phi\rangle\langle \delta\phi|]
\end{equation}
where $\rho = \int D\chi |\Psi\rangle\langle\Psi|$ is the density matrix obtained by tracing out $\chi$.  

%{\color{cyan} Mehrdad: I don't understand this paragraph} {\color{magenta} Eva:  I have edited to try to make it clearer, let me know if this addresses your question}.   

%In principle, we could work directly with the state functional $\Psi(\delta\phi(\x), \chi(\x), t_0)$, specifying it at time $t_0$ and evolving it back using the unitary evolution operator in the FRW geometry to reverse engineer the initial state.  
There is a wide range of initial conditions that lead to inflation; see \cite{Eastetal,Kleban:2016sqm}\ for some recent developments.   However, we will simply start from the Bunch-Davies vacuum.  This is a conservative choice for our purposes, as it avoids introducing non-Gaussianity at the level of the initial state.   We would like to understand the possible N! enhancement of $0\to N$ processes in the time dependent background, so we start in the vacuum, with no particles in the initial state.

To separate issues we will prescribe various time-dependent couplings which can be mediated by fields that evolve outside the horizon, e.g. at reheating, as discussed extensively in the early literature on multifield inflation and non-Gaussianity such as \cite{LocalNGmechanisms}.  In particular, we will introduce mixing between $\chi$ and $\delta\phi$ after they have evolved independently over $\sim N_e$ e-foldings.  

{

To begin, {for each mode $k$, there is a time $t_{c,k}\sim \log(k/k_*)/H$ at which it has just exited {the horizon}.  Let us denote by $t_c$ the time at which all modes accessible in the CMB have exited the horizon.}  At this time}, as just mentioned, we have a direct product state
\begin{equation}\label{productstate}
\Psi(\delta\phi,\chi, t_c) =\psi_G(\delta\phi, t_c)\psi_\perp (\chi, t_c)
\end{equation}
where we are neglecting slow-roll corrections and hence $\psi_G$ is the approximately Gaussian state of the inflaton fluctuations, of the form
\begin{equation}\label{psiGform}
\psi_G (f) \sim \sqrt{{\rm det}(C)} \exp(- f C{{}^{-1}} f)
\end{equation}
with covariance matrix
\begin{equation}\label{CGauss}
C \sim \delta(\k+\k') { {P_{\delta\phi}(k)}}, ~~~~~ P_{\delta\phi}(k)\sim \frac{H^2}{k^3}
\end{equation}
encoding scale invariant perturbations.  

There are in principle many choices for the state of the transverse sector and its dynamics.   We will consider a light field $\chi$, of mass $m_\chi\ll H$, starting in its ground state.  
For the full range of $\chi$ , we take its potential energy $V(\chi)$ to be subdominant to the inflaton potential in sourcing inflation; the slow roll conditions are satisfied separately in the $\chi$ directions.  
The interactions in the $\chi$ sector build up over a large number of e-foldings $N_e$, with each mode outside the horizon affected by a stochastic distribution of shorter modes \cite{Starobinsky}\cite{SalopekBond}.  In the next section, we will illustrate this buildup of nonlinearities.  In some cases we may focus on the late-time limit, and its equilibrium 1-point probability distribution.   Here each `point' is a patch of size the correlation length, $R_S$ described below, and the distribution obtained by tracing over the other patches is the equilibrium solution to the appropriate Fokker-Planck equation \cite{Starobinsky}\ 
\begin{equation}\label{latedist}
\int D\chi(x\ne x_0) |\Psi_\perp(\chi)|^2\to \rho_{eq}\sim {\cal N}_{eq} e^{- 4\pi^2 V(\chi (x_0))/3 H^4}
\end{equation}
where ${\cal N}_{eq}$ is a normalization factor.   This was worked out in detail, with a focus on the $\lambda\chi^4$ theory in \cite{Starobinsky}.  
Subleading corrections to this classical stochastic approximation and its derivation from the full quantum field theory were examined in \cite{GS}.  
%{\color{cyan} Is this correct? I think the corrections are either the out-of-equilibrium effects $\propto e^{-\sqrt{\lambda} N_e}$, or the $\O(\sqrt{\lambda})$ corrections to the equilibrium distribution.}

{

Similar results hold for multiple $\chi$ fields, and other potentials, but with an interesting subtlety.  
To explain what we mean by this, let us focus on potentials of the form 
\begin{equation}\label{Vpowers}
V(\chi)=\mu^{4-p}|\chi|^p
\end{equation}
This is a particular family of models motivated by the potential-flattening effects of multiple, generically massive, fields as we review further below \cite{flattening}.  The behavior at the origin in (\ref{Vpowers}) may be smoothed out by integrating in additional fields, but for the present discussion this will not be necessary and in fact the form (\ref{Vpowers}) leads to a very simple analysis.  

The Fokker-Planck equation for the one-point pdf of the long modes of $\chi$ takes the form
\begin{equation}\label{FPone}
\frac{\partial\rho_1}{\partial t}=\frac{H^3}{8\pi^2}\frac{\partial^2\rho_1}{\partial\chi^2}+\frac{1}{3 H}\frac{\partial}{\partial\chi}\left({V'(\chi)\rho_1}\right)
\end{equation}
The equilbrium solution arises from setting $\partial \rho_1/\partial t =0$.  To capture the approach to equilibrium (when it pertains), we need more general solutions.  
It is useful to work as reviewed in \cite{PBHpaper}\ in a basis of eigenstates of the operator on the right hand side, which gives an analogue Schrodinger problem \cite{Starobinsky}
\begin{equation}\label{StarSchrod}
\left( -\frac{\partial}{\partial\chi^2} + [v'(\chi)^2-v''(\chi)]\right)\Phi_n(\chi)=
\left(-\frac{\partial}{\partial\chi}+v(\chi)\right) \left(\frac{\partial}{\partial\chi}+v(\chi)\right)
\Phi_n(\chi)=\frac{8\pi^2\Lambda_n}{H^3}\Phi_n(\chi)
\end{equation}
with $v(\chi)=4\pi^2 V(\chi)/3 H^4$.  
The effective potential $w(\chi)\equiv  v'(\chi)^2-v''(\chi)$ in this problem leads to a vanishing lowest eigenvalue, $\Lambda_0=0$; this corresponds to the solution $\propto e^{-v(\chi)}$ (\ref{latedist}) as can be seen immediately from the middle form of (\ref{StarSchrod}).  When the nonzero eigenvalues $\Lambda_{n>0}$ are gapped, as we approach equilibrium the non-equilibrium terms are suppressed exponentially, $\sim e^{-\Lambda_n (t-t_0)}$.  

In the family of models (\ref{Vpowers}), the effective Schrodinger potential $w(\chi)$ has a delta function potential well at the origin which holds the ground state (or a smoothed version with $\Lambda_*$ turned on).  (This comes from the $v''(\chi)$ term, with the derivatives acting on the cusp at the origin.)   For $p > 1$, $w(\chi)\to\infty$ as $|\chi|\to\infty$ and the energy levels are discrete.  For $p=1$,  $w(\chi)$ approaches a positive constant at large field values:  there is a continuum above a gap, with $\Lambda_{gap}/H\sim (\mu/H)^6$.  
For $p<1$, $w(\chi)\to 0$ as $|\chi|\to\infty$, leading to an ungapped continuum of excited states.  It is straightforward to verify in this formalism that the $p=0$ case reproduces free field theory fluctuations.

%{\color{cyan} MM: I find this confusing. In the $p<1$ case there is a continuum of plane waves starting from $\Lambda =0$. Are we saying that there is also a bound state with zero energy? Isn't there a theorem that forbids degeneracy in 1d quantum mechanics? This would be in contrast to $\Phi_0$ given by \eqref{latedist}. It seems that here one should talk about quasi-normal modes of a metastable vacuum. } {\color{blue} ES: For all $p>0$, the ground state $\Phi_0$ is zero energy ($\Lambda_0=0$) and as Starabinsky-Yokoyama prove, the eigen-energies are in general non-negative; I introduced into the middle of our equation (\ref{StarSchrod}) the form from Starobinsky that makes this manifest.  The continuum of scattering states is not really at zero energy, it approaches zero from above as you take the wavefunction momentum (conjugate to $\chi$) to zero, but there is not really a constant solution strictly at zero.  I guess this means there is an order of limits, and I don't think this contradicts the theorem and it is very explicit.  (In any case, we have no business taking $\chi$ literally to infinity, so our scenario does not probe the nearly ungapped states.  Our plateau ends at some finite $\chi$ in all cases, if only because $V$ hits some UV scale where this description breaks down.)}

}

%As just stressed, this expression for the local field evolution includes the effects of the short modes.  These are controlled by an effective coupling {$\lambda N_e^\alpha$, for a real number $\alpha$ which depends on the model}.  
%{But one should not be tricked by the apparent simplicity of~(\ref{psichinext}): after a given mode has crossed the horizon, it evolves under the non-linear potential, but it is also affected by the fact that shorter modes keep coming out of the horizon, affecting the dynamics of the longer mode (as if they were changing its initial condition at that time). Here we just wish to stress that the evolution is local and classical, so that the final wavefunction for the field at a given  location is determined by a local-in-space deterministic classical evolution, averaged over the values that all the modes shorter than the mode of interest obtain at horizon crossing. }
%It would be interesting to determine explicitly the typical shape of this nonlinear function, given a particular model of the interactions $V(\chi)$.  Even for mildly perturbative couplings, this is nontrivial because of the stochastic effects.  But even without taking into account the stochastic effects, strong nonlinearities develop for $\lambda N_e\lesssim 1$, as we will see shortly in explicit examples.                 
%}

\subsection{Mixing with $\phi$ and the probability distribution for $\zeta$}

Finally at a late time $t_0$, to convert $\chi$ to $\delta\phi$, we introduce a mixing interaction
\begin{equation}
    \mathcal{S}_{mix} = \int dt \int d\mathbf{x}\, a(t)^3 F_{mix}(\chi) \dot{\phi}^2
\end{equation}
with support between times $t_0$ and $t_0+\Delta t$. We can understand the effect of this interaction by noting that during inflation, $\dot{\phi} = \dot{\overline{\phi}}(t) + \delta \dot{\phi}(\mathbf{x}, t)$, where the first term is the leading homogeneous piece. Thus, the interaction is, to leading order
\begin{equation}
    \mathcal{S}_{mix} \sim \int dt \int d\mathbf{x}\, \dot{\overline{\phi}} [a(t)^3 \delta \dot{\phi}] F_{mix}(\chi)
\end{equation}
We can write this in terms of the conjugate momentum to the inflaton fluctuation $\delta\phi$, $\Pi_{\delta\phi} = a(t)^3 \delta \dot{\phi}$ leading to a mixing Hamiltonian
\begin{equation}
    H_{mix} = \int d\mathbf{x}\, \dot{\overline{\phi}}\,\Pi_{\delta\phi} F_{mix}(\chi)
\end{equation}
that dominates over the free Hamiltonian, as described in  \cite{PBHpaper}; for completeness we briefly summarize the setup here.
%{\color{blue} In general there is a lot of repetition in the setup between this paper and the PBH one. Should we spell out all the details here too?} 
The operator $\Pi_{\delta\phi}$ is the generator of translation in field space and so the evolution over $\Delta t$ is just a shift of the wavefunction:
\begin{equation}
    \Psi(\chi, \delta\phi, t_0 + \Delta t) = \Psi(\chi, \delta\phi + \dot{\overline{\phi}}\Delta t F_{mix}(\chi), t_0)
\end{equation}

Putting all this together, the likelihood for $\delta\phi\sim \zeta \dot\phi/H$ is then given to good approximation by
\begin{equation}\label{mixed}
{\cal L}(\delta\phi|\{\lambda\}, \kappa) =\int D\chi_0 ~ |\psi_\perp (\chi_0, t_0)|^2 ~ |\Psi_G(\delta\phi +\kappa F_{mix}(\chi_0))|^2
\end{equation}
up to $1/N_e$ corrections.   Here we have defined $\kappa \equiv \dot{\overline{\phi}}\Delta t$.
%Here we have neglected a Jacobian in changing variables in the path integral; this is subdominant {in a logarithmic sense}. 
After this step of evolution, we postulate that reheating quickly leads to a local thermal distribution, with $\delta\phi\sim \zeta\dot\phi/H$ distributed according to the likelihood (\ref{mixed}).  Given this, $\zeta$ remains constant during the remaining evolution outside the horizon, and (\ref{mixed}) contains the primordial non-Gaussianity.  

For the case where we reach the equilibrium distribution in the $\chi$ sector, the one-point pdf for $\delta\phi$, defined as in (\ref{histfromPavg}), is easily computed by Gaussian integration
\begin{equation}\label{histapprox}
\langle N_{\delta\hat\phi}\rangle=\int d\vec\chi_0 \,{\cal N}_{eq} \exp(-4\pi^2 V(\vec\chi_0)/3 H^4) \frac{\exp(-(\delta\hat\phi{+}\kappa F_{mix}(\vec\chi_0))^2/2\sigma^2)}{\sqrt{2\pi}\sigma} \ ,
\end{equation}
where we used (\ref{latedist}), and we have allowed for the possibility of multiple $\chi$ fields.   Here the width $\sigma$ is given by
\begin{equation}\label{sighist}
\frac{1}{2\sigma^2}= C^{-1}_{x', x'}+ 4 C^{-1}_{x', \perp}(C^{-1}_{\perp,\perp})^{-1}C^{-1}_{\perp, x'} 
\end{equation}
where $C$ is the covariance matrix in position space, and $\perp$ denotes points not equal to $ x'$.  This width is of order $H$.  Again, (\ref{sighist}) can be traded for the $\hat\zeta\sim H\dph/\dot\phi$ histogram.

\subsection{Regime of applicability of the equilibrium distribution}\label{sec:equilibriumregime}
{
Let us now spell out the regime of applicability of the equilibrium distribution.  This depends in part on the relative size of various relevant patches.  

In the derivation of the equilibrium distribution, following \cite{Starobinsky}\ let us denote the correlation length as
\begin{equation}
R_S \sim H^{-1}e^{H/\Lambda_1}.
\end{equation}
As reviewed in \cite{PBHpaper}, this can be read off from the two point correlation function.  

%I agree with the rest of the argument if all $\epsilon$'s are replaced by $e^{-1/\sqrt{\lambda}}$ (or $e^{-H/\Lambda_1}$). 
%To have more than one patch in the observable universe, each of which larger than the resolution, we need
%\be
%e^{n_e-\Delta n_e} < R_S H < e^{n_e}
%\end{equation}
%which gives \eqref{parameterwindow}.}
%In the $\lambda\chi^4$ case, this is bounded by $1/\epsilon < e^{1/\sqrt{\lambda}}$, and we will note the generalization to other models in our family below.  
We must compare this to two other scales.  First, we have the size of the observable patch of the universe,
\begin{equation}\label{Robs}
R_{obs}=\frac{1}{H}e^{n_e}
\end{equation}
where $n_e<60$ is the total number of efoldings of phenomenological inflation.  The third scale of interest is the scale of resolution of the CMB, or of some shorter-scale probe such as PBHs.  This we will parameterize as
\begin{equation}\label{Rres}
R_{res}\sim \frac{1}{H} e^{n_e-\Delta n_e}
\end{equation}
The number of independent patches is 
\begin{equation}\label{Np}
N_P = \left(\frac{R_{obs}}{R_S}\right)^3 
\end{equation}
To have more than one patch in the observable universe, each of which larger than the resolution, we need
\begin{equation}
e^{n_e-\Delta n_e} < R_S H < e^{n_e}
\end{equation}

In other words, the equilibrium distribution applies in a straightforward way for 
\begin{equation}\label{parameterwindow}
\frac{1}{n_e}< \frac{\Lambda_1}{H} < \frac{1}{n_e-\Delta n_e}
\end{equation}
where the eigenvalues $\Lambda_n$ depend on the model parameters as determined by (\ref{StarSchrod}). For example, for the $p=1$ model $V(\chi) = \mu^3\chi$ we find $\frac{\Lambda_1}{H}\sim \frac{\mu^6}{H^6}$, while for the $p=4$ model $V(\chi)\sim \lambda \chi^4$ we have $\frac{\Lambda_1}{H}\sim \sqrt{\lambda}$ \cite{Starobinsky}.

If we restricted attention to the CMB, then this particular scenario, with the $\chi$ sector reaching the equilibrium distribution, pertains for a rather particular value of the coupling in this family of models (although one which might arise in a rich potential landscape).   Moreover, once it reaches equilibrium, the contribution $\chi$ makes to the fluctuations is very blue.   For shorter scale probes, such as primordial black holes, there is a wide window of applicability as described in \cite{PBHpaper}.  However, at least in that context the stochastic evolution of $\chi$ is only applicable to the leading observables if the potential drifts outward, for reasons explained in \cite{PBHpaper}.}  As we will review further below, the mixing interaction itself can introduce strong non-Gaussianity associated with the tail of the distribution.

\subsection{Flattened directions in field space and Non-Gaussian tails}

The effect of $\vec\chi$ on the histogram for $\dph$ can be understood analytically to some extent.   We will be particularly interested in the tails of the distribution.  To see whether or not the Gaussian tail dominates for $\dph\gg H$, consider field configurations where the Gaussian suppression is canceled by the $\vec\chi_0$ field: 
\begin{equation}\label{tailcanc}
F(\vec\chi_{0, tail})\simeq -\dph/\kappa.    
\end{equation}
In that regime, the probability is suppressed by $\exp(-4\pi^2 V(\vec\chi_{0, tail})/3 H^4)$.   If in this direction (or directions) in field space, the potential $V(\vec\chi_{0, tail})$ is flatter than quadratic in $\dph$, then the Non-Gaussian tail dominates over the Gaussian at sufficiently large $\dph$.  In order to be potentially observable, the overall probability of this tail must be larger than $1/N_{P}$ where $N_P$ is the number of independent data points in the survey volume: roughly,
\begin{equation}\label{Probcondition}
\int_{tail} d\dph \frac{{\cal N}_{eq}}{\sqrt{2\pi}}\exp(-4\pi^2 V(\vec\chi_{0, tail}(\dph))/3 H^4) > \frac{1}{N_{P}}
\end{equation}

Flattened potentials arise naturally from adjustments of heavier fields as in \cite{flattening}\ as well as for other reasons such as those studied in \cite{DBIStochastic}\cite{alpha}.  
In fact,  constraining the Non-Gaussian tail in our scenario gives us a new way to probe large field ranges, in the $\chi$ sector rather than the inflaton sector.  The less efficient our conversion is (i.e. for smaller mixing $\kappa$), the larger the field range is that we probe.\footnote{This is somewhat reminiscent of observations in \cite{LocalNGmechanisms}.}  Here, we probe the field range via the quantum (effectively stochastic) fluctuations of $\chi$ rather than the classical motion of $\chi$, and via non-Gaussianity rather than the tensor to scalar ratio.  

{ Of course, the distributions differ in other ways than asymptotically on the tail.  In some cases, including an axionic $\chi$ field, the Non-Gaussian histogram contains an intermediate region where it exceeds the Gaussian, before rejoining the Gaussian tail further out.  When probability moves to a region away from the origin in $\dph$, this is made up by a suppression of probabiliy near the origin.  Low-point moments are sensitive to the latter effect, and it is a quantitative question to determine which measurements best capture the difference in the two distributions.   PBH formation is directly sensitive to the tail, as we analyzed in \cite{PBHpaper}.  But other parts of the distribution may lead to other signals and constraints to take into account.  We will comment on this briefly after deriving the factorial enhancement in a wide class of multifield models.}

\section{The generating functional for connected $N$-point functions and $N!$ enhancement}\label{WJNpf}

One tractable probe of the distribution is its moments, the $N$-point correlation functions.  Also from a purely theoretical point of view, we would simply like to deteremine the fate of the factorial enhancement \cite{Voloshin}\ in our cosmological setting.  

We can estimate the $N$ dependence of the $N$-point functions by first extracting the connected ones by computing the generating functional
\begin{equation}\label{WJgen}
Z(J)=e^{-W[J]}=\int D\delphi\; {\cal L}[\delphi|\lambda, \kappa] \;e^{-\int J \delphi}
\end{equation}
with the connected $N$-point function given by
\begin{equation}\label{Wderivs}
{\left.\frac{\delta^N W}{\delta J_{\k_1}\dots \delta J_{\k_N}} \right|_{J_{\k_i}=0}} \sim \langle \delphi_{\k_1}\dots \delphi_{\k_N} \rangle_{C}.
\end{equation}
and the disconnected diagrams obtained from a similar formula with $W(J)$ replaced with $Z(J)$.  
For our case, the likelihood takes the special form (\ref{mixed}), so we get
\begin{equation}\label{WJus}
e^{-W[J]}=\int D\chi_0|\psi_\perp[\chi_0]|^2 \int D\delphi P_G[\delphi-\kappa F(\chi_0)] e^{-\int J \delphi}
\end{equation}
The path integral over $\delphi$ is a Gaussian, and gives 
\begin{equation}\label{WJnext}
e^{-W[J]}\sim e^{J { C} J} \int D\chi_0 |\psi_\perp[\chi_0]|^2 e^{-\int \kappa J F(\chi_0)}
\end{equation}
where $C$ is the Gaussian covariance (\ref{CGauss}).  We will discuss saddle point estimates for the $\chi_0$ integral below, working with the 1-point pdf (histogram).  

But first, we will derive $W(J)$ for a special choice of $\psi_\perp$ and $F(\chi)$ which is nontrivial but completely calculable in the full quantum field theory.  For the transverse state, we will simply take a Gaussian $\psi_G(\chi_0)$ (\ref{psiGform}).   For the mixing interaction, we consider
\begin{equation}\label{quadchi}
F(\chi_0)  = \frac{\chi_0^2}{M} + \chi_0
\end{equation}
This form arose from interesting (p)reheating dynamics in \cite{BondBilliards}.  
If the mass parameter $M$ is of order $H$, this is fully nonlinear;  the effective coupling is $H/M$.  
  
\subsection{Full field theory calculation in a special case}
    
For this case, the result is 
\begin{align}\label{WJquad}
W[J] &\sim \int d\k d\k' \sqrt{P_\delphi(k)}J_\k { \left[ \delta_{\k,\k'}+\kappa^2\left(\delta_{\k\k'}+\kappa \frac{J_{-\k+\k'}}{M}\sqrt{P_\delphi(k)P_\delphi(k')}\right)^{-1} \right]}\sqrt{P_\delphi(k')}J_{\k'} \nonumber\\
&  ~~~~~~ -\frac{1}{2} {\rm Tr~ log} \left({\delta_{\k \k'}} + \kappa { \frac{J_{-\k+\k'}}{M}}\sqrt{P_\delphi(k)P_\delphi(k')} \right) + \text{const} \nonumber \\
% &\sim& J C^{-1}J -\frac{1}{2} Tr \log\left(\frac{\delta_{\k \k'}}{P(k)} + \kappa J_{-\k-\k'} \right)  \nonumber \\
%&\sim&  J C^{-1}J -\frac{1}{2} Tr \sum_{n=1}^\infty \frac{(-1)^{n+1}}{n}\left(\kappa \sqrt{P P} J \right)^n + const \nonumber \\
\end{align}

%\begin{align}\label{WJquad}
%W[J] &\sim& J C^{-1}J -\log det \left(\frac{\delta_{\k \k'}}{P(k)} + \kappa J_{-\k-\k'} \right)^{1/2}  \nonumber \\
% &\sim& J C^{-1}J -\frac{1}{2} Tr \log\left(\frac{\delta_{\k \k'}}{P(k)} + \kappa J_{-\k-\k'} \right)  \nonumber \\
%&\sim&  J C^{-1}J -\frac{1}{2} Tr \sum_{n=1}^\infty \frac{(-1)^{n+1}}{n}\left(\kappa \sqrt{P P} J \right)^n + const \nonumber \\
%\end{align}

%In the last expression, we expanded the log.  The matrix appearing in each term is $B^n$ with  $B_{\k_1\k_2}= \kappa\sqrt{ P(k_1)P(k_2)} J_{-\k_1-\k_2}$, and matrix multiplication being integration/summation over indices $\k$.   

Evaluating the derivatives (\ref{Wderivs}) after expanding $W[J]$ in a power series in $J$ gives us the following result for the $N$-point function.  From the top line of (\ref{WJquad}) we obtain, { for $N>2$,}~\footnote{The $N$-point functions for $\zeta$ are obtained by the rescaling $\zeta\sim H \delta\phi/\dot\phi$.}
\begin{align}\label{Npftree}
\langle\delta\phi^N\rangle|_0 &\sim \frac{\kappa^N }{M^{N-2}}\delta(\sum\k) P_\delphi(k_1)P_\delphi(|\k_1+\k_2|) P_\delphi(|\k_1+\k_2+\k_3|)\dots P_\delphi(|\k_1+\dots +\k_{N-2}|) P_\delphi(k_N)\nonumber\\ &+ {\rm permutations} \nonumber\\ &\sim \kappa^N N!
\end{align}
which has the structure of a tree diagram.  
From the second line we obtain
\begin{align}\label{Npfloop}
\langle\delta\phi^N\rangle|_1&\sim\frac{\kappa^N }{M^N} \delta(\sum\k)\int d^3\k P_\delphi(k)P_\delphi(|\k_1+\k|)P_\delphi(|\k_2+\k_1+\k|)\dots P_\delphi(|\k_{N-1}+\dots +\k_1+\k|)  \nonumber\\
&+ {\rm permutations} \nonumber\\ 
&\sim \kappa^N N! 
\end{align}
This has the structure of a loop diagram.  
These contributions both have the expected scaling with momenta for a nearly scale-invariant theory.
The amplitude is enhanced by $N!$.  {The overall level of non-Gaussianity of the map is naively of order $\frac{\langle\delta\phi(x)^N\rangle_c}{\langle\delta\phi(x)^2\rangle^{N/2}}\sim N! \kappa^N \left(\frac{H}{M}\right)^{N-2}$ {from (\ref{Npftree})}}, but this does not generally reflect the actual observable level.  

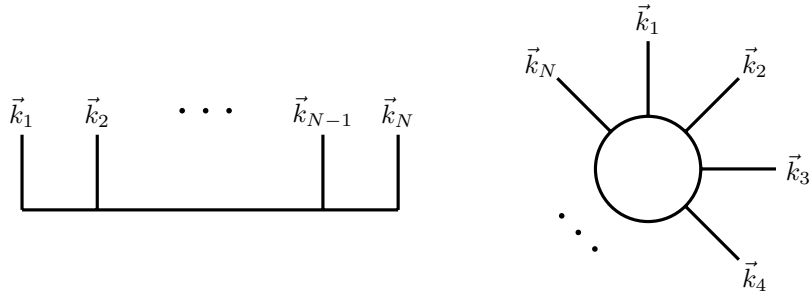
\begin{figure}[htbp]
  \centering
  \begin{subfigure}[c]{0.49\textwidth}
    \centering
      \begin{tikzpicture}
        \draw [very thick]
        (0,0) -- (5,0)
        (0,0) -- (0,1)
        (1,0) -- (1,1)
        (4,0) -- (4,1)
        (5,0) -- (5,1);
        \node at (0,1.3) {$\vec{k}_1$};
        \node at (1,1.3) {$\vec{k}_2$};
        \node at (4,1.3) {$\vec{k}_{N-1}$};
        \node at (5,1.3) {$\vec{k}_N$};
        \node [scale=2] at (2.5,1.3) {$\dotsm$};
      \end{tikzpicture}
    \end{subfigure}
  \begin{subfigure}[c]{0.49\textwidth}
    \centering
      \begin{tikzpicture}
        \draw [very thick]
        (0,0) circle (0.7)
        (0,0.7) -- (0,1.7)
        (0.7*0.71, 0.7*0.71) -- (1.7*0.71, 1.7*0.71)
        (0.7,0) -- (1.7, 0)
        (0.7*0.71, -0.7*0.71) -- (1.7*0.71, -1.7*0.71)
        (-0.7*0.71, 0.7*0.71) -- (-1.7*0.71, 1.7*0.71);
        \node at (0,2) {$\vec{k}_1$};
        \node at (2*0.71, 2*0.71) {$\vec{k}_2$};
        \node at (2, 0) {$\vec{k}_3$};
        \node at (2*0.71, -2*0.71) {$\vec{k}_4$};
        \node at (-2*0.71, 2*0.71) {$\vec{k}_N$};
        \node [rotate=-45,scale=2] at (-0.9, -0.9) {$\dotsm$};
      \end{tikzpicture}
    \end{subfigure}
\caption{A diagrammatic representation of the two contributions to the $N$-point function described in the text, (\ref{Npftree}) on the left and (\ref{Npfloop}) on the right.}
\label{diagramsNpf}
\end{figure}

%The $N$-point function (\ref{Wderivs}) is given by N functional derivatives of the $n=N$ term in the power series expansion of $W[J]$.  It has $N!$ terms from permutations of where these derivatives act, and the coefficients in the expansion are not factorially suppressed.  As a result, the amplitude of the $N$-point function scales like $\kappa^N N!$.     

For more general $\psi_\perp(\chi)$ and $F(\chi)$, we can obtain similar results, now using a saddle point approximation to the integral.  Shortly we will see that for a rather generic (but not entire) nonlinear function, the order $N$ term in the expansion of $W[J]$ has no factorial suppression. Hence, $N$-point correlators obtained by Nth functional derivative will be factorially enhanced. To make this clear, we can study a 1d integral version of the problem, the histogram of scalar fluctuations defined above.  We will show this in the next subsection.  

%Then, given the surviving factorial enhancement of the $N$-point functions at large N, we will be in a position to compare their signal/noise and constraining and discovery potential with that of the bispectrum.   

\subsection{More general theories and the factorial enhancement}

We would like to understand how general the factorial enhancement is given a more generic model than the one just analyzed.  Clearly small perturbations around the example above will not change the conclusion.  More generally, we can analyze this by considering the histogram version of the integral (\ref{WJnext}).  This suffices to capture the N-point functions at coincident points, and hence it is enough to determine the factorial structure.  (However, it does not necessarily capture the strongest tails in the full quantum field theory.)   For this exercise, let us consider the histogram arising from the equilibrium distribution \cite{Starobinsky}\ discussed above:    
\begin{equation}\label{WJnextoy}
Z(J)=e^{-W(J)} \sim \int d\vec\chi_0 \exp\left(-\frac{4\pi^2 V(\vec\chi_0)}{3 H^4} - \kappa J F(\vec\chi_0) \right).
\end{equation}
We have not included the $e^{JC^{-1}J}$ term here as it only contributes to the 2-point function, or the normalization since this drops out of the N point function growth. 
%In this integral we have essentially downgraded $J$, $\chi_c$ and the covariance matrix to ordinary numbers. 
The $N$-point functions are obtained by taking N ordinary derivatives of $W(J)$, which still captures the combinatorial factors.  These depend on the behavior of the coefficients in a series expansion of $W(J)$ (or $Z(J)$ for the disconnected diagrams). 

We can assess the combinatorial factor using the structure of the integrals that arise in the expansion with respect to $J$.  The disconnected diagrams are generated by 
\begin{equation}\label{Zexpansion}
Z(J) = \sum_n z_n \kappa^n J^n   
\end{equation}
The corresponding disconnected N point functions go like $z_N N!\kappa^N$.  So these have a factorial enhancement if the coefficients $z_n$ are not suppressed by $1/n!$, in which case the series has a finite radius of convergence (possibly zero, meaning the series is only asymptotic).  For that to be the case, the function $Z(J)$ should not be entire.  One way that a function can fail to be entire is if it diverges somewhere in the complex $\kappa$ plane.  This in turn depends on whether the potential $V(\vec\chi_0)$ grows more quickly than $F(\vec\chi_0)$ in every direction in field space.  If not, then the disconnected diagrams have a factorial growth, and the distribution has a non-Gaussian tail at sufficiently large $\dph$, as discussed above.     

For the connected N point functions derived from $W(J)$ we have a similar criterion for a factorial enhancement.  This may have an enhancement even when the disconnected correlators do not (one can see an example of this simply from the fact that taking the logarithm leads to non-analyticity at the zeros of (\ref{WJnextoy})).    

We proceed by evaluating the integral by saddle point. The saddle point equation for $\chi_{0I}^*$ is:
\begin{equation}\label{saddleqn}
\frac{4\pi^2 \partial_IV(\vec\chi_0^*)}{3 H^4} + \kappa J \partial_I F(\vec\chi_{0}^*) = 0
\end{equation}
It will be useful to express the function $F$ evaluated on the solution to this equation by a power series
\begin{equation}\label{saddleqnsol}
F(\vec\chi_{0}^{*}(J)) = \sum_{n=0}^\infty a_{n} J^n
\end{equation}
The saddle point value of $W(J)$ is
\begin{equation}\label{Wchicstar}
W(J) =\frac{4\pi^2 V(\vec\chi_0^*)}{3 H^4} + \kappa J  F(\vec\chi_{0}^*).
\end{equation} 
Differentiating $W(J)$ with respect to $J$ and using (\ref{saddleqn}) we obtain the differential equation
\begin{equation}\label{Wdiffeqn}
\frac{dW}{dJ} = \kappa F(\vec\chi_{0}^{*}(J))
\end{equation}
which upon integration gives
\begin{equation}\label{Wpowseries}
W(J) = \kappa \sum_{n=0}^\infty \frac{a_n J^{n+1}}{n+1}
\end{equation}
{after we impose that $W(J=0)=0$.}
The $N$-point functions are thus $\sim a_N N!$. The main question is then how the coefficients $a_N$ scale. If they cancel the $N!$ enhancement then clearly the power expansion of $F(\vec\chi_0^{*}(J))$ must converge for all complex J and thus the function must be entire. This is a very stringent requirement and is not true in general. 

To simplify the analysis, let us now specialize to the case of a single variable $\chi_0$.  Equation (\ref{saddleqn}) can be viewed as an inversion problem. We are given $J = g(\chi_0^*)$, where the function $g$ is  
\begin{equation}\label{gdef}
g(x)= -\frac{4\pi^2 V'(x)}{3 H^4 \kappa F'(x)}
\end{equation}
and we want to show that $F\circ g^{-1}$ is not entire. Consider, for example, $F(x)\propto x$. Then, a sufficient (but not necessary) condition is that $g'$ has finite roots in the complex plane as that would imply that the derivative of $g^{-1}$ has a pole at that point.
Furthermore, for this particle example for $F$, this corresponds to the question of whether $V''(x)$ has roots in the complex plane.  

%classical nonlinear evolution due to a quartic potential (\ref{chicchi0four}). The saddle point equation is then
%\begin{equation}label{chicchi0foursaddle}
%27 H^6 \kappa^2 \sigma_c^4 J^2 = \chi_c^{*2} (3H^2 + 2 \lambda_4 N_e \chi_c^{*2})^3
%\end{equation}
%The right hand side of this equation is analogous to the function $f$ discussed above. Its first derivative clearly has roots and thus $\chi_c^{*2}(J)$ cannot be entire.

\subsection{Example of non-analytic $W(J)$}
It is possible for the saddle point equation to have a solution which is not Taylor expandable. A simple but important example is a potential of the form $V(\chi_0) = |\frac{\chi_0}{M}|^p$, with $p > 1$ and $F(\chi_0) = \chi_0$. The integral is then (neglecting all constants as they can be absorbed in the definition of $J$)
\begin{equation}\label{WJnonanalytic}
    Z(J) = e^{-W(J)} \sim \int_{-\infty}^\infty d\chi_0 \exp\left(-|\chi_0|^p - J\chi_0\right) = 2\int_0^\infty dx e^{-x^p}\cosh(Jx)
\end{equation}
To conclude that the connected correlation functions exhibit an $N!$ enhancement, it suffices to show that there exists a $J_0\in \mathbb{C}$ such that $Z(J_0) = 0$. Then $W$ has a { logarithimic branch cut} at $J = J_0$ and is thus not analytic everywhere.

To that end, let $J$ be purely imaginary and define $J\equiv i y$. Then,
\begin{equation}
    Z(iy) = 2\int_0^\infty dx e^{-x^p} \cos(y x)
\end{equation}
This as a function of $y$ is purely real and continuous. Therefore, if we can prove that it takes a negative value, it would imply that it also has a zero. For concreteness, take $y=2\pi$
\begin{equation}
    Z(2\pi i) = 2\int_0^\infty dx e^{-x^p} (1-2\sin^2(\pi x)) = 2\Gamma\left(1+\frac{1}{p}\right) - 4\int_0^\infty dx e^{-x^p} \sin^2(\pi x)
\end{equation}
We can make a series of approximations to the final integral
\begin{align}
    \int_0^\infty dx e^{-x^p} \sin^2(\pi x) &> \int_0^1 dx e^{-x^p} \sin^2(\pi x) > \int_0^1 dx (1-x^p)\sin^2(\pi x) \nonumber\\
    &> \frac{1}{2} - \int_0^1 dx x^p \pi^2 (x-1)^2 = \frac{1}{2} - \frac{2\pi^2}{(p+1)(p+2)(p+3)}
\end{align}
to conclude that
\begin{equation}
    Z(2\pi i) < 2\Gamma\left(1+\frac{1}{p}\right) - 2 + \frac{8\pi^2}{(p+1)(p+2)(p+3)}
\end{equation}
which is negative for $p > 6.4$ and goes as $-\frac{2\gamma}{p}$ for large $p$.

For values close to $p=2$, we can numerically Taylor-expand $Z(2\pi i)$ around $p=2$. This gives
\begin{equation}
    Z(2\pi i) \approx 9.17\times 10^{-5} - 3.85\times 10^{-2} (p-2)
\end{equation}
which is negative for $p >  2.003$. We can fill in the intermediate regime by taking more terms in the approximations above. The result is that $Z(2\pi i)$ is negative for all $p > 2.003$.

\section{Comments on observational implications}

It is interesting to apply these results to primordial non-Gaussianity searches.  It sharpens the question of systematically mapping out the ideal probe of Non-Gaussianity (low point functions versus the histogram or higher moments).\footnote{As mentioned above, the dominance of higher point functions has arisen previously in examples \cite{BondBilliards}\cite{AndreiWeb}\cite{PBHpaper}\cite{PajerN}\cite{productive}\cite{Bruno}. Another previous incarnation of this question led to a negative result in a different context as explained in \cite{34confusion}.} In the present context, this may be model-dependent as a result of the exponential dependence of the tail of the distribution on the fields and parameters.

In \cite{PBHpaper}\ we focused on primordial black hole production, which occurs on shorter scales than the CMB.  In this section, we will consider the histogram (\ref{histfromPavg}) which might be applied to the CMB or large scale structure.  As described above in section \ref{sec:equilibriumregime}, the applicability of the stochastic nonlinearities is limited to a narrow (but nonvanishing) window in coupling (\ref{parameterwindow}).  However, the mixing itself introduces heavy tails of the distribution in appropriate cases, and in those examples there is no such limitation.    

\subsection{Signal to Gaussian noise formula and its limitations}\label{sec:SNprobs}

In the collider physics version of this quantum field theory problem \cite{Voloshin}\cite{Browntree} \cite{Argyresetal} \cite{Sonetal} \cite{Khozeetal}\cite{criticalrefs}, the quantity of physical interest is the cross section (squared N point function amplitude).  This is factorially enhanced at tree level, sufficiently close to threshold.  The analogous squared quantity in our case, formally, would be signal to noise estimate for an N point function estimator.

In all examples with a factorial enhancement, the ratio of the non-Gaussian mean and the Gaussian variance, which we review shortly, is similarly factorially enhanced. This by itself would naively indicate a generic new discovery window for non-Gaussianity.  However, it is necessary to analyze the full distribution of the estimator to determine how likely such a discovery would be, and this turns out to be model-dependent.   

By working in the cosmic variance limited regime of CMB observations, we can focus on the noise introduced by the 
quantum fluctuations of the fields themselves.  In general, this is highly nontrivial, with a covariance matrix 
\begin{equation}\label{Ncovar}
C^{(N)}_{\{\k_1,\dots,\k_N\}, \{\k'_1,\dots,\k'_N\}}=\langle\zeta_{\k_1}\dots\zeta_{\k_N} \zeta_{\k'_1}\dots\zeta_{\k'_N}\rangle
\end{equation}  
which is a 2$N$-point function.  

Including only the noise from Gaussian fluctuations, and including only connected contributions to the $N$-point functions, this matrix is diagonal and leads to a relatively simple expression
\begin{eqnarray}\label{SNsq}
&&(S/N)^2 =\int_{\{\k\}\,,\{\k'\}}\langle\zeta_1\dots\zeta_N\rangle_{C}^{ *}\; {C^{(N)}}{(\{\k\},\{\k'\})}^{-1}\; \langle\zeta_{1'}\dots\zeta_{N'}\rangle_{C}\\ \nonumber
&&\qquad \to\quad  \int_{\{ \k \}} \frac{|\langle\zeta_{\k_1}\dots\zeta_{\k_N}\rangle_C|^2}{N! \prod P(k_i)} \equiv (S/N)^2_G
\end{eqnarray}    
where 
\begin{equation}\label{Pzeta}
P (k) \sim \frac{H^4}{\dot\phi^2 k^3} 
\end{equation}
is the power spectrum for $\zeta$.
Here the $N!$ in the denominator compensates for the unrestricted momentum integrals over the $N$ identical fields in the final state 
%{\cyan I think we are somewhat squeezing the notion of quantum to call these $N$ quanta. What about identical fields?}. 
This is similar
to the $1/N!$ arising in the multiparticle density of states for scattering with identical final particles.  The integrals over phase space are restricted to
\begin{equation}\label{kminkmax}
k_{\rm min}< \{ |{\bf{k}}|\} < k_{\rm max}  
\end{equation}
where  $k_{\rm min}\sim 1/L$ with $L$ the size of the survey, and $k_{\rm max}$ is the largest momentum scale we can probe.  One can analyze this quantity, finding that it has an interesting enhancement related to the N! growth of correlators.  Nonetheless, the probabilility of a detection for a given $N_{pix}$ is model-dependent within this class.
The reason that the nominal S/N is not a good guide is that the distribution of the estimator may be highly non-Gaussian.   We see that explicitly below in figure \ref{fig:likelihoods}.          

\subsection{Basic estimates of observational sensitivity}

One diagnostic of the information available to distinguish the Non-Gaussian probability distribution from the Gaussian one is the relative entropy (also known as the Kullback–-Leibler divergence), an average of the log of the ratio of likelihoods at two values of some theoretical parameter $\lambda$:
\begin{equation}\label{RelSdef}
{\cal S}_{rel}\equiv \int D\zeta P(\zeta)\log\left(\frac{{\cal L}(\zeta(\x) |\{\lambda\})}{{\cal L}(\zeta(\x) |\{0\})}\right)
\end{equation}
Here, $P(\zeta)$ may be taken to be either of the two probability distributions; ${\cal S}_{rel}$ is not symmetric.  The first term in its Taylor expansion 
about $\{\lambda\}=0$ is the Fisher metric $F_{\lambda\lambda}$.  As we will see, in some cases, the relative entropy (\ref{RelSdef}) is well approximated by the first term in the Taylor expansion,  the constraint on $\lambda$ is well estimated by the inverse of the Fisher metric, and low-point correlation functions suffice to achieve this constraint.  In other cases, this first term is subdominant, and there is more information available (e.g. on the tail of the distribution).   Moreover, certain observables (such as primordial black hole production \cite{PBHpaper}) are specifically sensitive to the tail.   

{The analysis above establishes factorial enhancement of N point functions for the families of models described above in (\ref{powersVf}), and it is clear that this extends to many others. 
We note that the factorial enhancement of the connected diagrams is universal in this class, while that of the disconnected diagrams is model dependent.
For example, we can 
parameterize a class of models by
\begin{equation}\label{powersVf}
V = \mu^{4-p} (\Lambda_*^2+\chi^2)^{p/2}, ~~~~~~~~~~ F(\chi)=  H \left(\frac{\chi}{H}\right)^m \xrightarrow[m \to \infty]{} H e^{2{\chi}/M_*}
\end{equation}
The tail becomes stronger with larger $m/p$.  Small values of $p$ emerge from the flattening mechanism discussed in \cite{flattening};  moreover, with more generic kinetic terms, the possibilities proliferate, at least in some cases leading to a flatter distribution for different reasons \cite{DBIStochastic}\cite{alpha}.  Large integer values of $m$ do not appear particularly well-motivated a priori, but the $m\to\infty$ limit leads to a Wilsonian-natural model of a hyperbolic field space 
\begin{equation}\label{hypmodel}
F(\chi) \sim  H e^{2{\chi}/M_*}
\end{equation}
similar to the structure of the kinetic terms considered in e.g. \cite{hyperbolic}.  As mentioned in \cite{PBHpaper}, this   
has a very heavy tail compared to the Gaussian case.  We can think of the first expression in (\ref{powersVf}) as an ad hoc parameterization of the slope of the potential in the direction giving the strongest contribution to the tail.  
In this class, the heavier than Gaussian tails only arise for $p<m$.   So for example, the $\chi^4$ theory with mixing $m\le 4$ has a Gaussian tail asymptotically, but still has factorial-enhanced connected N point functions.}  One can also analyze fields with an underlying periodicity, something also considered in \cite{Bruno}.  In the case (\ref{hypmodel}), the dominant contribution to the non-Gaussianity is from the mixing interaction, liberating us from the condition (\ref{parameterwindow}) as anticipated above.

% As we will see analytically below, all examples in this class have a factorial enhancement in the connected N point functions, but the strength of the disconnected N point functions and the tail will depend on the flatness of $V(\vec\chi_0)$.  

%\bea\label{mixed}
%{\cal L}(\delta\phi|\{\lambda\}, \kappa) &=&\int D\chi_0 ~ |\psi_\perp (\chi_0, t_0)|^2 ~ |\Psi_G(\delta\phi +\kappa \chi_0)|^2 \nonumber\\
%&\simeq& \int D\chi_c P_G(\delta\phi{+}\kappa \chi_0(\chi_c)) P_G(\chi_c) \ ,
%\end{equation}a
{
%\color{magenta}

%Secondly, we will consider a periodic direction in field space, with a negligible potential; 
%\begin{equation}\label{circlecase}
%V(\chi_0)\ll H^4, ~~~~~~ \chi_0 = %\chi_0+2\pi f_a
%\end{equation}
%and a conversion factor that respects the periodicity $F(\chi_0)=\sin\chi_0/f_a$.  

%On the other hand, it is interesting to consider the phenomenology for a pure $\chi^4$ potential, the $p=4$ case.  For example, the Standard Model of particle physics and its mixing to the inflaton sector -- if the inflaton is not the Higgs -- may be constrained.  We will address this example specifically below.     

%The correction in the denominator of (\ref{chicchi0}) could be large consistently with the slow-roll approximation (\ref{SRchicond}) because of this factor.  

{
%\color{green} 

\subsubsection{Corrections to the power spectrum ($N=2$)}

Before considering the tail of the distribution, it is interesting to ask what the effect of the mixing is on the power spectrum.  
At order $\kappa^2$, we get a correction to the $2$ point function.     
First, we note that
\begin{equation}\label{1pf}
\langle\delta\phi\rangle = \kappa \langle F(\chi_0)\rangle + {\cal O}(\kappa^3)
\end{equation}
Let us shift away the unobservable zero mode, defining 
\begin{equation}\label{shiftphi}
\delta\phi = f +\langle\delta\phi\rangle
\end{equation}          
where $\langle \delta\phi \rangle \simeq \kappa \langle\chi_0\rangle$.  
We then have a probability distribution
\begin{equation}\label{Pf}
{\cal L}(f|\kappa)= \int D\chi_0 |\psi_\perp[\chi_0]|^2P_G[f+\langle\delta\phi\rangle -\kappa F(\chi_0) ],
\end{equation}
the likelihood of measuring a fluctuation $f$ given $\kappa$.

Let us define $P_{\chi_0}(k)$ by 
\begin{equation}\label{deftwo}
\langle F(\chi_{0})_{\k_1}F(\chi_{0})_{\k_2} \rangle= \int D\chi_0 |\psi_\perp[\chi_0]|^2  F(\chi_{0})_{\k_1}F(\chi_{0})_{\k_2} \equiv P_{\chi_0}(k_1)\delta(\k_1+\k_2).
\end{equation}
Expanding the likelihood in $\kappa$, we find
\begin{equation}\label{Pfexp}
{\cal L}(f|\kappa)= {\cal L}(f|0)\left(1+ \frac{1}{2} \kappa^2 \int d\k \frac{P_{\chi_0}(k)}{P_{\delta\phi}(k)^2}f_\k f_{-\k} +\dots\right)
\end{equation}
with 
\begin{equation}\label{Pf0}
{\cal L}(f| 0) = \exp\left( -\frac{1}{2}\int d\k \frac{1}{P_{\delta\phi}(k)}f_\k f_{-\k} \right)
\end{equation}
At order $\kappa^2$, this simply means
\begin{equation}\label{powershift}
P_{\delta\phi}(k)\to P_{\delta\phi}(k)+\kappa^2 P_{\chi_0}(k)
\end{equation}
The $\chi$ sector modifies the power spectrum at order $\kappa^2$.  

In the regime we are focused on, with couplings satisfying $\lambda n_e^2\ge 1$,
%$\lambda N_e\sim 1${}~\footnote{\color{magenta} Here and in the rest of the paper, $\lambda\sim 1/N_e$ will be shorthand for $\lambda\sim 1/N_e^\alpha$, where the leading non-linear correction from outside of the horizon evolution goes as $\lambda N_e^\alpha$. }, 
the function $P_{\chi_0}(k)$ will have a fully nonlinear dependence on $\log(k)$. 
%{\color{cyan} Shouldn't this be $\lambda n_2^2>1$?}
In other words,  it will not be a simple perturbative expansion in tilt, running, etc., in contrast to minimal single-field slow roll inflationary models.  In the absence of non-Gaussianity, this could potentially provide an upper bound on $\kappa$ of order
\begin{equation}\label{kapupper}
\Delta\kappa|_{2pf}\sim  \frac{1}{N_{P}^{1/4}}\sqrt{\frac{P_{\delta\phi} }{P_{\chi_0}}}
\end{equation}
which in itself is an improvement over the bound from the bispectrum constraint on $f_{\rm NL}^{\rm local}$, which scales like $ N_{\rm p}^{-1/6}$ in this regime.  Conversely, there is a similar improvement in the discovery potential in the two point function given (\ref{kapupper}). 

}}

\subsubsection{Information in the tail for a family of models}

Here we analyze the Non-Gaussian histogram quantitatively for the family of models defined in (\ref{powersVf}).  Although the factorial enhancement of connected N point functions is general, the accessible information beyond the 2 point function is model-dependent.  We will classify the regimes according to the behavior of the histogram and the various N point functions.  (Even the 2 point function is informative, especially for theories with { $\lambda n_e^2>1$ }, as there
is no suppression of the running versus the tilt and so on.)  

\smallskip

{\bf{Analytic estimates for the size of the tail}}

\smallskip

Before getting into detailed analysis, we can estimate the size of the tail at the upper bound on $\kappa$ that could be inferred from a bound on corrections to the 2-point function.  Let us consider the class of models described above (\ref{powersVf}).  For these, we can write the histogram as 
\begin{eqnarray}\label{powersdist}
\langle N_{\delta\hat\phi}\rangle &=& \int d\chi_0 \,{\cal N}_{eq} \exp(-4\pi^2 \mu^{4-p}|\chi_0|^p/3 H^4) \frac{\exp(-(\delta\hat\phi{+}\kappa H (\chi_0/H)^m)^2/2\sigma^2)}{\sqrt{2\pi}\sigma} \ , \nonumber\\
&=&  \int d\tilde\chi_0 \,\tilde{\cal N}_{eq} \exp(-|\tilde\chi_0|^p) \frac{\exp(-(\delta\hat\phi{+}\tilde\kappa H \tilde\chi_0^m)^2/2\sigma^2)}{\sqrt{2\pi}\sigma} \nonumber\\
\end{eqnarray}
where the only parameter that enters is 
\begin{equation}\label{kaptilde}
\tilde\kappa = \frac{\kappa}{(\mu (4\pi^2/3)^{1/(4-p)}/H)^{m(4-p)/p}}
\end{equation}
To estimate the size of the tail, we use the relations
\begin{equation}\label{tailrelations}
\tilde\chi_{0, tail}^m\sim \frac{\dph/H}{\tilde\kappa} \sim \frac{\tilde\chi_{0, tail}^{p/2}}{\tilde\kappa}
\end{equation}
The first relation here is (\ref{tailcanc}), and the second is the crossover between the dominance of the Gaussian in $\dph$ and the dominance of the tail, $\sim\exp(-\tilde\chi^p)$.    Putting these together, we have a suppression of the tail by a factor
\begin{equation}\label{tailkappa}
\exp(-\frac{1}{\tilde\kappa^{p/(m-p/2)}})
\end{equation} 

In this section, we will imagine that we have observational access to all $N$ point functions, and work out the information content of the tail versus low point correlators.    In \cite{PBHpaper}\ we focused on an application to PBH formation, which is specifically sensitive to the tail (although even in that context, the variance can play a role as in \cite{SMBHmu}).
                        
In that spirit, if we evaluate $\tilde\kappa$ at the bound it is possible to obtain from the 2 point function
\begin{equation}\label{tilddap2pf}
\tilde\kappa^2 \frac{\Gamma(\frac{1+2m}{p})}{\Gamma(\frac{1}{p})} < \frac{1}{\sqrt{N_{P}}}
\end{equation}
this scales like 
\begin{equation}\label{tailscale}
\exp\{-N_{P}^{p/(4(m-p/2))}\left(\frac{\Gamma(\frac{1+2m}{p})}{\Gamma(\frac{1}{p})}\right)^{p/(2m-p)} \}
\end{equation}

For the special model described around (\ref{quadchi}), we effectively have
$p=2, m=2$.  (In this case, we are not working with the equilbrium Starobinsky distribution, but the model is equivalent to the one with these values of $p$ and $m$.) With $N_P=N_{pix}\sim 10^6$, this evaluates to $\exp(-\sqrt{N_{pix}}\Gamma(5/2)/\Gamma(1/2))\simeq 10^{-326}$, hence nowhere near observable in the CMB.    But relatively small changes in parameters make a big difference; larger $m$ (e.g. of order 10) leads to much less suppression.  Formally, smaller values of $p$ would also do this, but dialing $p$ in that way introduces the need to satisfy (\ref{parameterwindow}).  

For the analysis in this section, we will illustrate the information content by
considering different ratios of $p/m$.  This captures the effect of dialing up the parameter $m$, which is motivated by the fact that large $m$ matches onto the natural model (\ref{hypmodel}) on a hyperbolic field space geometry.      

%With $m=3$, the maximum of this function is $\gtrsim 1/N_{pix}$, at $p\sim .7$.  

 % (a value in the regime of UV models of effective single-field flat directions \cite{flattening}, in fact close to a particular power (2/3) that can arise via an interplay of axions and fluxes \cite{powers}).  
\smallskip

{\bf{Numerical analysis}}

\smallskip

Here, we construct realizations of the Non-Gaussian distributions discussed in the above section. {We evaluate whether low point correlation functions are in principle best for detecting them, or whether instead other aspects of the distribution such as the tail or higher point correlators contain more information.  }

For the purposes of this section, we will use the following family of distributions:
\begin{equation}
 P(-\infty < \phi < \infty) = \frac{1}{2\sqrt{2\pi}\Gamma\left(\frac{1}{p}+1\right)}\int_{-\infty}^\infty d\chi \exp\left(-|\chi|^p - 
\frac{(\phi - {k}^{\frac{1}{p}} \chi^m)^2}{2} \right) 
\end{equation}
The relation between $k$ and $\tilde\kappa\propto k^{1/p}$ can be read off from (\ref{powersdist}) above.   With this normalized distribution and a numerical analysis, we will check the estimates made above for models with accessible information on the tail.

If we focus for simplicity on N-point functions, we can determine which $N$ would be best
for detecting the non-Gaussianity.  We generate a large number of Gaussian and non-Gaussian realizations (data sets), each containing $N_{P}$ points.  We evaluate the even N-point function estimator on each simulated map.  
For each $N$, we find the range in which 90\% of the Gaussian results fall, starting from zero.  In other words, for each correlation function, we find where the 90th percentile lies in the Gaussian realizations. We then compute the
percentage of Non-Gaussian realizations that are above that 90th percentile.  For tail-dominated models, such as the hyperbolic model (\ref{hypmodel}), this can be a large percentage as we will see in an example below.  
%In general, the Non-Gaussian result could be below the Gaussian, but that would imply that the tail is not larger than a Gaussian which is not the situation that we are trying to analyze here. 
We also do this for the likelihood.

To be more specific, we consider the following estimators for the N-point functions:
\begin{equation}
\hat{\mathcal{E}}_N = \frac{1}{N_{P}}\sum_{i=1}^{N_{P}} \phi_i^N
\end{equation}
where $\phi_i$ are the $N_{P}$ data points drawn from either a Gaussian or a Non-Gaussian distribution. We can also define an estimator by evaluating the log-likelihood on the map as follows:
\begin{equation}
\hat{\mathcal{E}}_L = 
%\frac{1}{N_{pix}}
\sum_{i=1}^{N_{P}} \log\left(\frac{P_{NG}(\phi_i)}{P_G(\phi_i)}\right)
\end{equation}
The relative entropies are just the expectation values of this estimator over the two distributions:
\begin{equation}
\langle S \rangle_{NG} \equiv E[\hat{\mathcal{E}}_L]_{NG} = \int dx P_{NG}(x) \log\left(\frac{P_{NG}(x)}{P_G(x)}\right)
\end{equation}
\begin{equation}
\langle S \rangle_{G} \equiv -E[\hat{\mathcal{E}}_L]_G = -\int dx P_{G}(x) \log\left(\frac{P_{NG}(x)}{P_G(x)}\right)
\end{equation}

\begin{figure}
  \centering
  \includegraphics[width=0.6\textwidth]{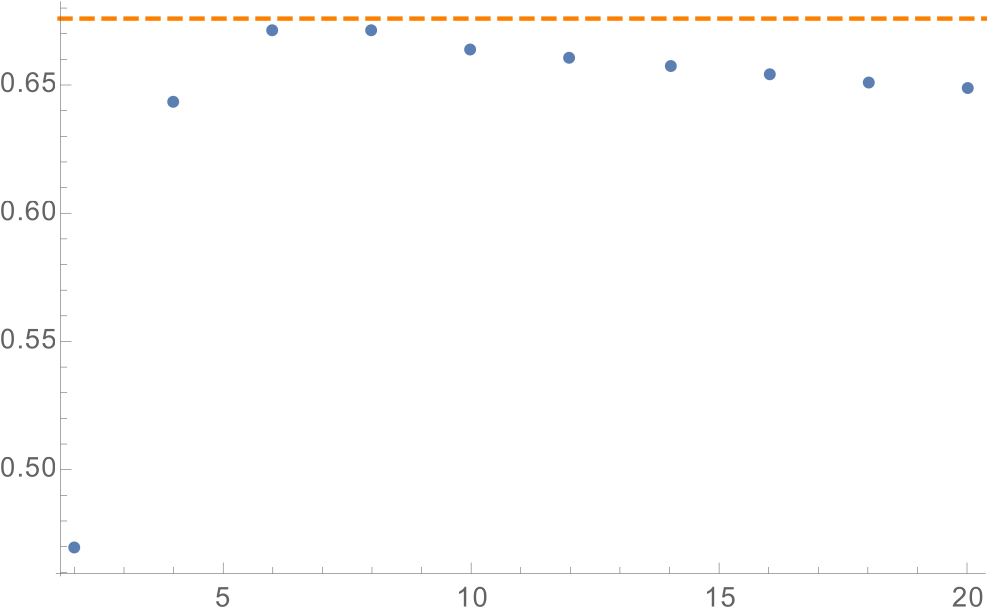}
  \caption{The horizontal axis is the even N-pt functions up to $N=20$ for the distribution with $m=3$ and $p=0.7$, with $N_{P}=1000$.  The vertical axis is the fraction of samples that are above the 90th percentile of those in the Gaussian distribution as described in the text.  The dashed line is the likelihood. The low point correlators are not optimal in this example. }
  \label{fig:nptfns}
\end{figure}

If we  Taylor expand around $\kappa=0$ in our distributions, the first surviving term is of order $N_{P}\kappa^4$, matching the two point function constraint.  One can compare this to the full relative entropy, computed with respect to either the Gaussian or non-Gaussian probability.  If these do not agree, then this indicates that the 2 point correlation function does not contain all the information.

%\be
%P_{NG}(x,k) = P_G(x) + k \frac{\partial}{\partial k} P_{NG} + \frac{k^2}{2} \frac{\partial^2}{\partial k^2} P_{NG} + \mathcal{O}(k^3)
%\ee
%and then the entropies become:
%\be
%\langle S \rangle_{NG} = \langle S \rangle_{G} = \int dx \frac{k^2}{2 P_G(x)}\left(\frac{\partial}{\partial k} P_{NG}\right)^2 + \mathcal{O}(k^3)
%\ee
%Thus, if the distribution is Taylor-expandable, we expect the two entropies to match. If they do not, that hints to a distribution with significant tail.}

As was discussed in the above section, the dominance of the tail is very sensitive to model parameters.  As one particular example, we expect the distribution with $m=3$ and $p=0.7$ to be tail dominated. 
Figure \ref{fig:nptfns} shows the results of the first 10 even N-pt functions for that particular distribution, with $k=1/6$.
The dashed line is the result of using the likelihood as our observable.
Clearly, the 2-pt function does not do a good job of detecting the Non-Gaussianity. The optimal N-pt functions are the 6-th and the 8-th in this case. { Conversely, in a non-tail dominated model, the 2-pt function would essentially be lying on the likelihood line with the successive N-pt functions decreasing and plateauing for large N.}
%{\color{cyan} Is there a way to determine the optimum $N$? Probably the one that converges at the onset of the tail.}

\begin{figure}
  \centering
  \begin{subfigure}[b]{0.3\textwidth}
    \includegraphics[width=\textwidth]{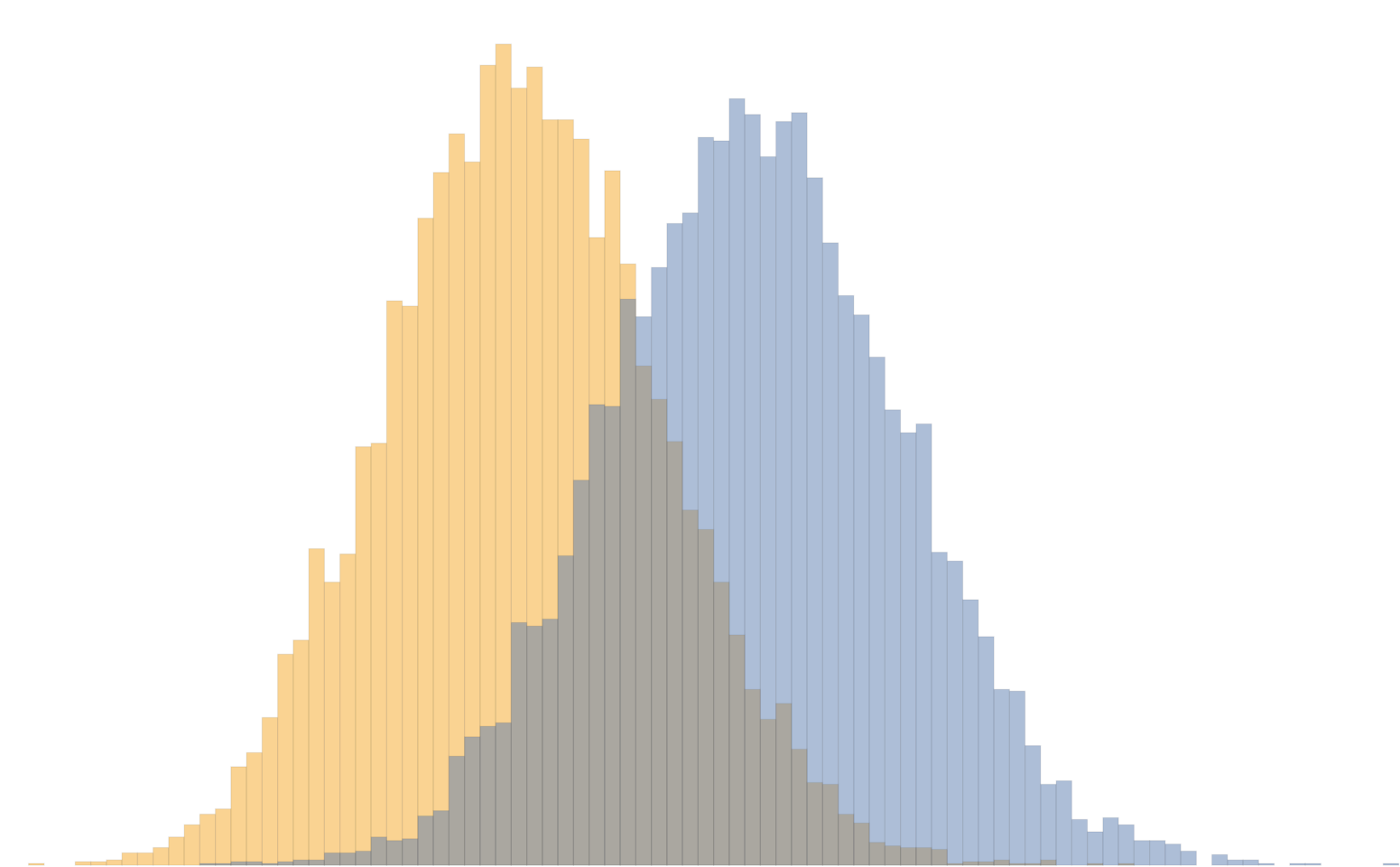}
    \caption*{likelihood}
  \end{subfigure}
  \begin{subfigure}[b]{0.3\textwidth}
    \includegraphics[width=\textwidth]{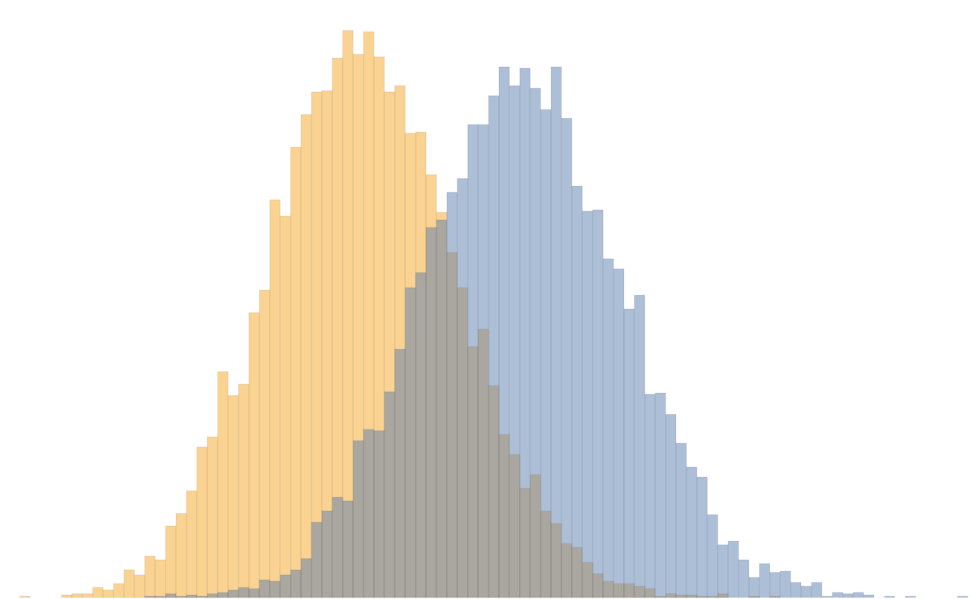}
    \caption*{2-point function}
  \end{subfigure}
  \begin{subfigure}[b]{0.3\textwidth}
    \includegraphics[width=\textwidth]{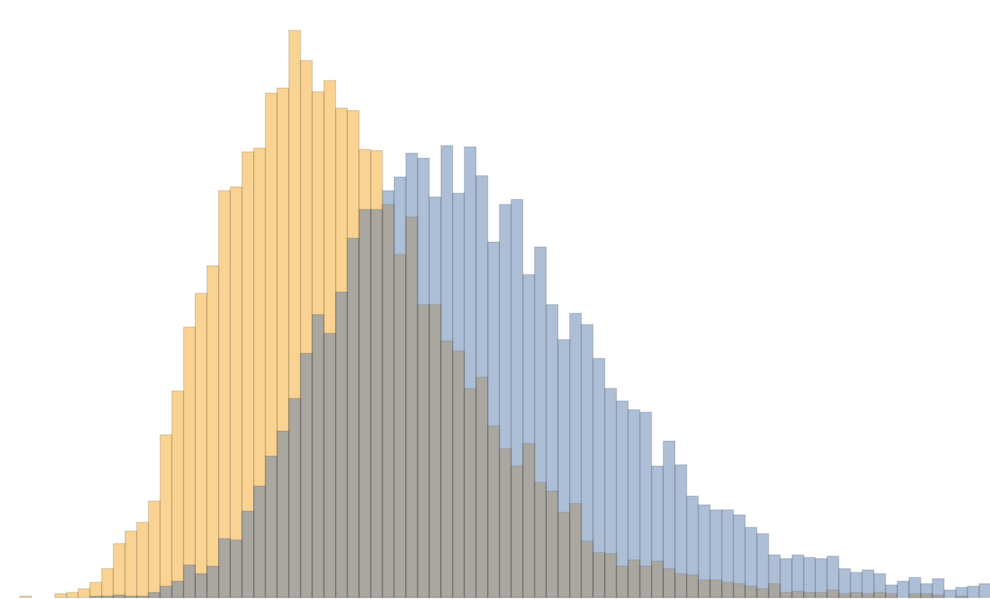}
    \caption*{6-point function}
  \end{subfigure}
  
  \begin{subfigure}[b]{0.3\textwidth}
    \includegraphics[width=\textwidth]{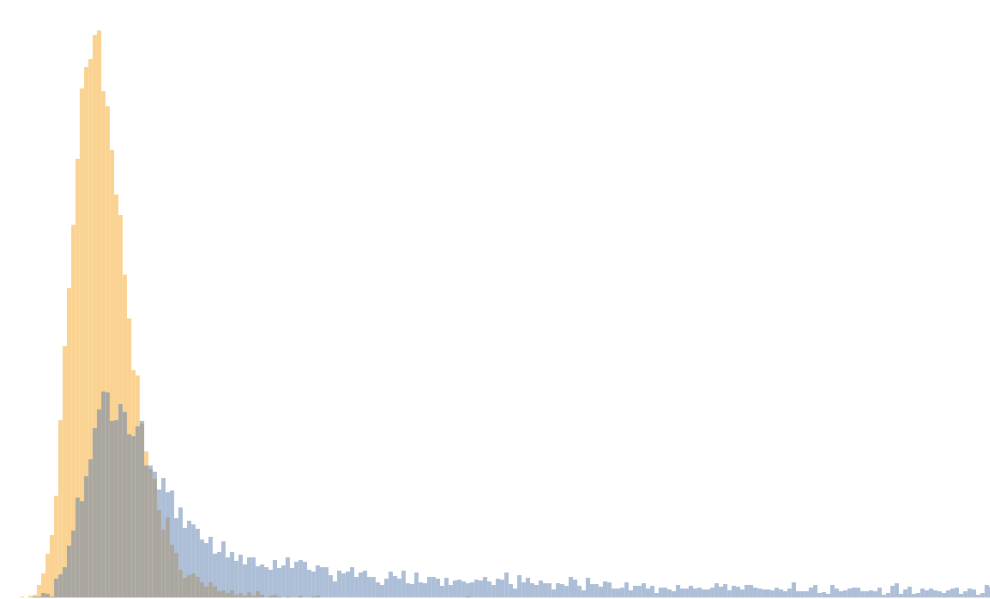}
    \caption*{likelihood}
  \end{subfigure}
  \begin{subfigure}[b]{0.3\textwidth}
    \includegraphics[width=\textwidth]{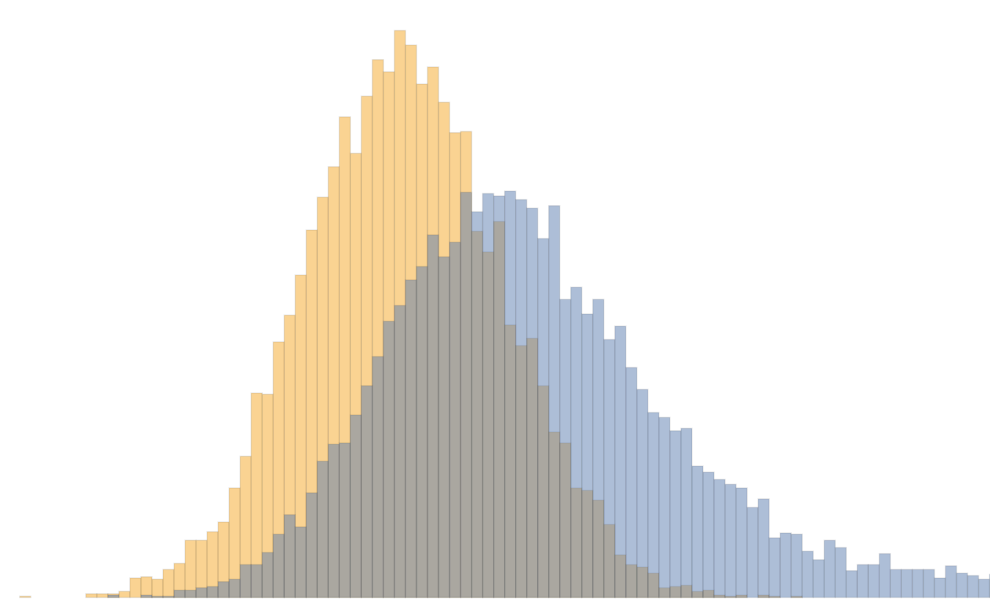}
    \caption*{2-point function}
  \end{subfigure}
  \begin{subfigure}[b]{0.3\textwidth}
    \includegraphics[width=\textwidth]{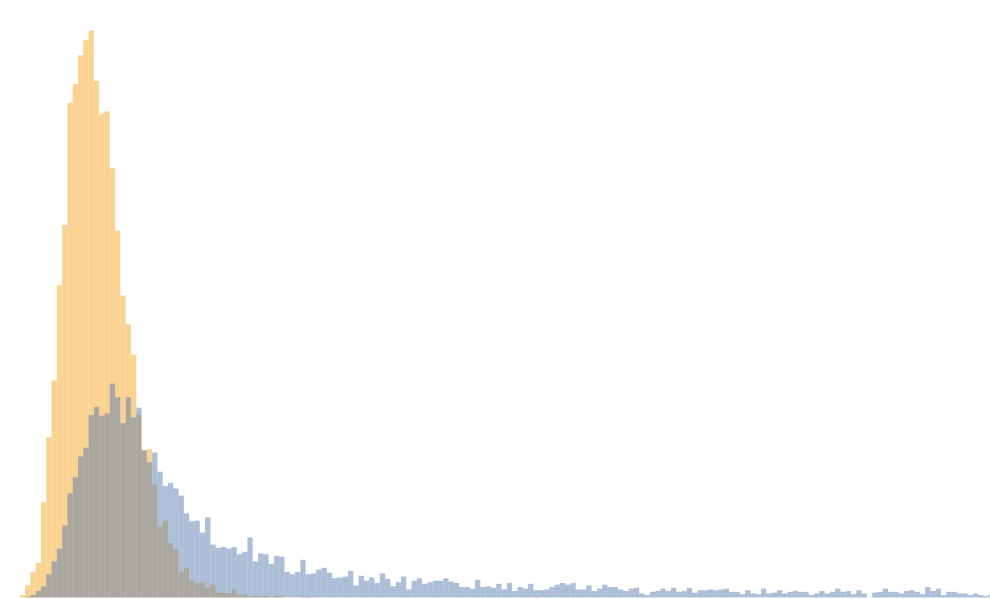}
    \caption*{6-point function}
  \end{subfigure}
  \caption{The likelihood and the distributions of the 2-point function and 6-point function estimators, for the Gaussian distribution (in yellow), and a non-Gaussian distribution (in blue). The first row is the model with $m=1$, $p=4$, $k=1/20$, $N_{P}=1000$ and the second row with $m=3$, $p = 0.7$, $k=1/6$, $N_{P}=1000$.}
  \label{fig:likelihoods}
\end{figure}

In Figure \ref{fig:likelihoods} the distribution of the likelihood, 2-pt function and 6-pt function are plotted. 
For the case with $m=1$ and $p=4$,
the 2-pt function distribution is essentially the same as the likelihood, while the 6-pt function has a significantly more tailed distribution.  This non-Gaussian distribution of the estimator illustrates why the naive signal/noise analyzed in section \ref{sec:SNprobs}\ -- which is generically factorial enhanced -- is not by itself an indicator of observational sensitivity.   
However, in the $m=3$ and $p=0.7$ model, the 2-pt function is very different from the likelihood.
This behavior is model-dependent, but applies to interesting models such as (\ref{hypmodel}).

Our results for different parameters are summarized in Table \ref{table:modelnums}. The results for models with $m\neq 1$ and $p$ are
similar to those with $m=1$ and $p^{\prime} = \frac{p}{m}$, and a large ratio of $m/p$ should 
be a good guide to the natural hyperbolic model (\ref{hypmodel}) \cite{PBHpaper}.  
%so we only give the ones with $m=1$ for brevity. {\cyan There is one with $m=3$.}
\begin{table}
  \begin{center}
    \begin{tabular}{c | c | c || c | c | c}
      m & p & k & $\langle S \rangle_G$ & $\langle S \rangle_{NG}$ & Best N-pf \\\hline
      1 & $1/16$ & $1/29$ & 0.86 & 73 & 8-12 \\
      1 & $1/8$ & $1/17$ & 0.86 & 35 & 6-10 \\
      1 & $1/4$ & $1/10$ & 1.1 & 5.8 & 4-6 \\
      1 & $1/2$ & $1/6$ & 1.9 & 2.5 & 2-4 \\
      1 & $1$ & $1/5$ & 1.4 & 1.5 & 2 \\
      1 & $4$ & $1/20$ & 1.3 & 1.4 & 2 \\\hline
      3 & $0.7$ & $1/6$ & 1.0 & 13.7 & 6-8
    \end{tabular}
  \end{center}
  \caption{Numerical results for different distributions with $N_{P} = 1000$ and $10000$ samples. $S$ is the relative entropy, computed either with respect to the Gaussian or Non-Gaussian distribution as indicated in the columns, including the factor of $N_{P}$. We chose this to be order 1, i.e. a barely detectable difference between the two distributions, according to the Gaussian-weighted relative entropy. {For tail-dominated models, we find a discrepancy between the two relative entropies, with a large relative entropy weighted with the non-Gaussian distribution.}}
  \label{table:modelnums}
\end{table}

\section{Conclusions and future directions}
In this work, we showed that in the multifield inflationary context, factorial enhancement of $N$ point correlation functions survives quantum effects and applies in the regime of kinematic interest.  
This is a basic question in quantum field theory motivated by the factorial enhancement known in particular parameter and kinematic regimes.  It is simpler to analyze more fully and exploit in the regime of physical interest in our cosmological setting than in collider physics
(although in that context this question has stimulated a number of interesting results \cite{Voloshin}\cite{Browntree} \cite{Argyresetal} \cite{Sonetal} \cite{Khozeetal}\cite{criticalrefs}).  The basic reason for this is the dilution of gradients, along with the calculably stochastic behavior of the system that applies in some regimes of couplings.  

Specifically, we derived and applied the enhanced amplitude of these large $N$-point functions in the study of primordial non-Gaussianity.  We encountered some subtleties along the way, but were left with interesting model-dependent possibilities for substantially enhanced sensitivity beyond low point correlators. It would be interesting to explore in more depth the phenomenological implications,
beyond that of enhanced primordial black hole production addressed recently in \cite{PBHpaper}.
%However, there could be other ways for this effect to be detectable. 

The models we analyzed in this work contain additional fields during inflation, which is reasonable given the multiple fields in the Standard model as well as hidden sectors that often arise in string theory.  This enabled us to apply the theory of stochastic inflation for certain windows of couplings,  as well as mixing interactions among field sectors in all cases. A natural question that arises is whether this effect persists in the case when any additional fields are too heavy during inflation to have such effects, reducing the system effectively to a single-field model of the primordial perturbations. We leave these questions to future work, perhaps building from recent progress on the calculation of multipoint correlators in other areas of quantum field theory \cite{criticalrefs}.  In the present work, the kinematic simplicity and resulting calculability of the ultralocal multifield dynamics in early universe inflation enabled us to settle the factorial enhancement question in the affirmative in this context.  

\bigskip

\noindent{\bf Acknowledgments}
We thank Mehrdad Mirbabayi, Leonardo Senatore, and Matias Zaldarriaga for extensive discussions and collaboration on this subject. We also thank N. Arkani-Hamed, J. R. Bond, J. Cardy, P. Creminelli, R. Flauger, V. Gorbenko, Z. Komargodski, M. Munchmeyer, D. Murli, J. Polchinski, S. Shenker,  K. Smith, D. Spergel, J. Thompson, and B. Wandelt for useful related discussions.
This research was supported in part by the Simons Foundation Origins of the Universe Initiative (modern inflationary cosmology collaboration), and by a Simons Investigator award.   GP and ES are grateful to the KITP for hospitality during part of this project.

\end{document}